\documentclass[aip,jcp,numerical,reprint]{revtex4-1}

\usepackage{graphicx}
\usepackage{amsmath}
\usepackage{amssymb}

\begin{document}
\title{Thermoelectric Performance of various Benzo-difuran Wires}

\author{Csaba G. P\'eterfalvi}
\affiliation{Department of Physics,
Lancaster University,
Lancaster, LA1 4YB, UK}

\author{Iain Grace}
\affiliation{Department of Physics,
Lancaster University,
Lancaster, LA1 4YB, UK}

\author{D\'avid Zs. Manrique}
\affiliation{Department of Physics,
Lancaster University,
Lancaster, LA1 4YB, UK}

\author{Colin J. Lambert}
\email{c.lambert@lancaster.ac.uk}
\affiliation{Department of Physics,
Lancaster University,
Lancaster, LA1 4YB, UK}

\newcommand{\igr}[2][]{\includegraphics[#1]{#2}}

\date{\today}

\begin{abstract}

Using a first principles approach to electron transport, we calculate the electrical and thermoelectrical transport properties of a series of molecular wires containing benzo-difuran subunits. We demonstrate that the side groups introduce Fano resonances, the energy of which is changing with the electronegativity of selected atoms in it. We also study the relative effect of single, double or triple bonds along the molecular backbone and find that single bonds yield the highest thermopower, approximately 22$\mu$V/K at room temperature, which is comparable with the highest measured values for single-molecule thermopower reported to date.

\end{abstract}
\pacs{31.15Ap, 31.15.Ei, 31.15.Xi, 31.15.aq, 31.50.Df, 73.23.Ad, 73.23.Hk, 73.63.Rt}

\maketitle

\section{Introduction}

The field of molecular electronics has been driven by the goal of producing electronic components at the sub 10\,nm scale. The advantage of using molecules to achieve this goal lies in the ability to control functionality through chemical synthesis. In this work we study the theoretical transport properties of a series of molecular wires containing benzo-difuran (BDF) units which have recently been synthesized as target components in molecular devices~\cite{Yi2010}. BDF derivatives have previously been targets for use in device applications such as organic light-emitting diodes, organic field-effect transistors, and photovoltaic cells, partly due to their hole transporting properties\cite{hole}. Here we consider a series of molecules with a BDF core and different substituents, which extend the \hbox{$\pi$-conjugation} of the BDF skeleton. These molecules are attractive for single-molecule electronics, because they form a class of highly conjugated heterocyclic molecules which open up the possibility of added functionality, including organic fluorescence\cite{Fluor1,Fluor2}. In this work we will focus on their electrical and thermoelectrical transport properties to assess their suitability as active components in molecular electronic devices.

\begin{figure}[h]
\centering
  \igr[width=62mm]{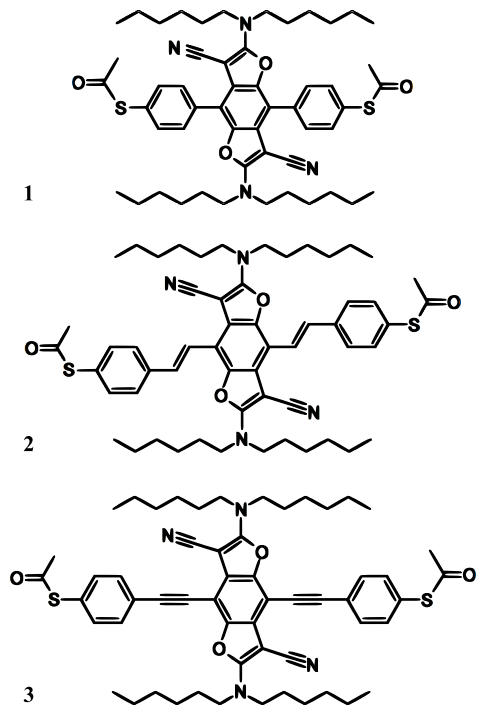}
  \caption{Structure of single, double and triple-bonded BDF molecules with thiol anchors and protecting groups}
  \label{fgr:BDFstructure}
\end{figure}

The recently-synthesized series of molecules containing BDF cores\cite{Yi2010} is shown in Fig.~\ref{fgr:BDFstructure}. These vary in the bonding between the three rings of the central backbone, with molecule {\bf 1} possessing the shortest backbone containing rings linked by a single carbon bond and molecule {\bf 3} possessing the longest with rings linked by triple bonds. Figure~\ref{fgr:BDFstructure} shows the series with thiol end groups together with the protecting groups. We are interested in the possibility that the furan side groups of these molecules generate Fano resonances in transport properties, since these can lead to enhanced thermoelectrical performance, provided they appear near the Fermi energy of the gold electrodes\cite{FanoPRB}.

\section{Theoretical Method}\label{sec:theory}

\begin{figure}[h]
 \centering
   \igr[width=86mm]{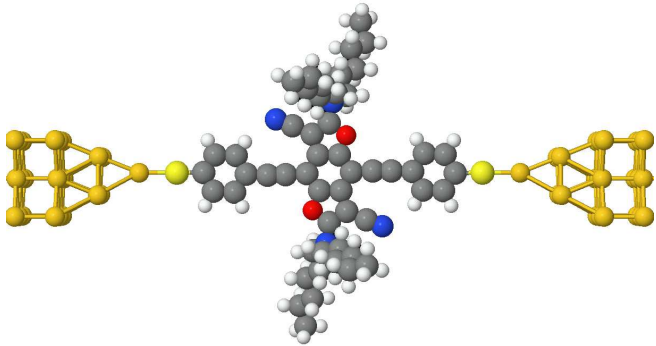}
   \caption{Structure of the extended molecule {\bf 3}. Thiol anchor groups bind to top atom of the gold surface.}
   \label{fgr:ExtMol}
\end{figure}

To calculate the conductance of this series of molecular wires, we use the {\it ab-initio} transport code SMEAGOL\cite{SMEAGOL,SMEAGOL2}. As we are interested in the low-voltage properties of these molecules, we use an 'equilibrium' implementation which more efficiently calculates the zero-bias transmission $T(E)$. The first step is to calculate the geometry of the isolated molecular wire, for which we use the density functional (DFT) code SIESTA\cite{SIESTA}. A double-zeta polarized (DZP) basis is chosen, the exchange correlation is described by generalized gradient approximation (GGA)\cite{GGA} and the atoms are relaxed until all forces are less than 0.02\,eV/{\AA}. An extended molecule is then created by adding the first few layers of gold to the ends of the molecule (see Fig.~\ref{fgr:ExtMol}). A 'pyramid' of gold atoms at the surface is chosen to mimic a typical break junction and the lead cross-section is taken to be 3 by 3 layers of (111) gold. Six layers of gold are included, which is enough to converge the transmission. The contacting geometry is taken to be the terminal sulphur binding to the top atom on the gold pyramid. The optimum binding geometry is found using DFT to minimize the gold-sulphur distance, which we find to be 2.5\,{\AA}.

A tight binding Hamiltonian describing this structure is then used to calculate the zero-bias transmission $T(E)$, where $E$ is the energy of the incoming electron and the conductance is obtained via the Landauer formula \hbox{$G(E) = G_0\ T(E)$}, where $G_0=2 e^2/h$.

The results produced using this method usually do not agree with experimental measurements due to inherent problems associated with DFT, which inaccurately predicts the HOMO and LUMO eigenvalues. Consequently the resonances  in $T(E)$ can appear at the wrong energies. In the literature, various corrections\cite{Quek2007,SAINT} have been implemented to overcome this limitation. Here we present an alternative method which corrects both the HOMO-LUMO gap and also the energies of the HOMO-1 and LUMO+1 states. The method involves comparing the above DFT-based $T(E)$ with the transmission $T_M(E)$ of a model Hamiltonian $H_M$ and then adjusting the parameters of the model such that model $T_M(E)$ coincides with DFT-based $T(E)$. More precisely, we adjust the model Hamiltonian parameters to minimize $\chi^2$, where $$\chi^2 = \int^{E_2}_{E_1}\left(\log T(E)-\log T_M(E)\right)^2\ .$$ The choice of $E_1$ and $E_2$ is somewhat arbitrary and chosen to define an energy window containing two resonances below and two other above the Fermi energy. In practice, the fit does not depend strongly on this choice, provided there are no strong resonances too close to the limiting values on either side. Just like the DFT-based $T(E)$ the resonances of the model $T_M(E)$ will appear at the wrong energies. However, since the model Hamiltonian is known, it is now a simple matter to adjust the model Hamiltonian parameters such that the resonances of $T_M(E)$ appear at the correct energies.

\begin{figure}[h]
\centering
  \igr[width=86mm]{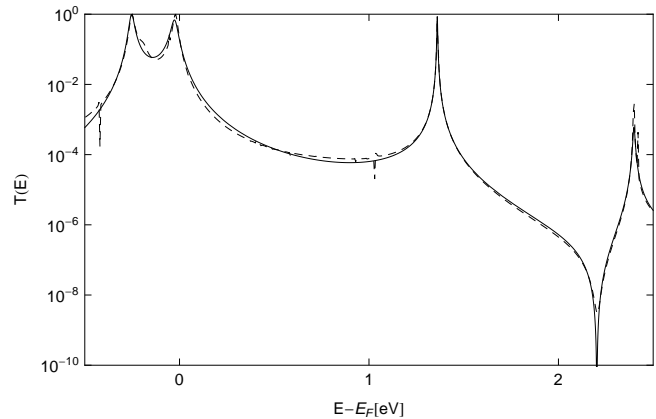}
  \caption{The original DFT-based transmission function $T(E)$ (dashed line) and $T_M(E)$, the uncorrected transmission probability of the model Hamiltonian (solid line) for molecule {\bf 3}. The average $\chi$ per data point is 0.129, 0.069 and 0.146 for molecules {\bf 1}, {\bf 2} and {\bf 3} respectively.}
  \label{fgr:corr}
\end{figure}

Since we are interested in correcting the position of four frontier resonances (HOMO, LUMO, HOMO-1 and LUMO+1), we choose a four-state model Hamiltonian of the form:
\begin{equation*}
H_M=\left(
\begin{array}{cccccc}
 0 & \gamma_{11} & \gamma_{21} & \gamma_{31} & \gamma_{41} & 0 \\
 \gamma_{11} & \epsilon_{1} & \kappa_{12} & \kappa_{13} & \kappa_{14} & \gamma_{12} \\
 \gamma_{21} & \kappa_{12} & \epsilon_{2} & \kappa_{23} & \kappa_{24} & \gamma_{22} \\
 \gamma_{31} & \kappa_{13} & \kappa_{23} & \epsilon_{3} & \kappa_{34} & \gamma_{32} \\
 \gamma_{41} & \kappa_{14} & \kappa_{24} & \kappa_{34} & \epsilon_{4} & \gamma_{42} \\
 0 & \gamma_{12} & \gamma_{22} & \gamma_{32} & \gamma_{42} & 0 \\
\end{array}
\right),
\end{equation*}
where $\epsilon_i$ is the energy of the state $i$, $\kappa_{ij}$ describes the coupling between the molecular states $i$ and $j$, and $\gamma_{il}$ stands for the coupling between the molecular state $i$ and the lead $l$, which can be the left one or right one. By decimating the Hamiltonian $H_M$ and applying Dyson's equation, we obtain the surface Green's function $G$ and the transmission probability $T_M(E)=|\nu G_{12}|^2$, where $G_{12}$ is the off-diagonal element of $G$ and $\nu$ is proportional to the group velocity in the identical leads\cite{Sparks}. This presents a fitting problem to the original DFT-based transmission function $T(E)$ with 18 free variables. The physical meaning of these variables is of course defined only up to a unitary transformation. To assign a physical meaning, one should diagonalise the $4\times 4$ central part of $H_M$. The resulting Hamiltonian then has 4 eigenvalues on the diagonal, 4 couplings to the left lead and 4 couplings to the right lead. The physical meaning of these 12 parameters is therefore clear. The resulting fitted matrix elements of $H_M$ can now be adjusted to yield resonances at the correct energies. Here we use wide-band approximation, in which $\nu$ is assumed to be energy independent\cite{Verzijl2013}. For molecule {\bf 3} shown in Fig.~\ref{fgr:ExtMol}, a comparison between $T(E)$ and the uncorrected $T_M(E)$ is shown in Fig.~\ref{fgr:corr}. The corrected $T_M(E)$ for the same molecule can be seen in Fig.~\ref{fgr:Thiol}.

In principle, the correct resonance energies could be obtained experimentally. In the absence of experimental values, we obtain more accurate values for the frontier orbital energies using time-dependent DFT (TDDFT). In what follows, we use the numerical software NWChem\cite{NWChem} with B3LYP\cite{B3LYP} exchange-correlation potential and \mbox{6-311G*} basis functions. The overlap integrals show that the lowest singlet excitation $S_0$ corresponds to the HOMO $\rightarrow$ LUMO transition. Similar statements can be made with next few excitations and the HOMO $\rightarrow$ LUMO+1 and HOMO-1 $\rightarrow$ LUMO transitions. Using the first singlet excitation energy yields a good estimate for the positions of the resonant peaks around the Fermi energy, because this is the minimum energy required for the formation of a free electron--hole pair\cite{Nayak2009}. In particular, when the Fermi energy is approximately in the middle of the HOMO-LUMO gap, the HOMO and the LUMO levels play equally important roles in the transport. Therefore the difference between their associated eigenenergies should be calculated when they are populated equally by one electron on each, which is given by the first singlet excitation. Two other excitations, corresponding to the HOMO $\rightarrow$ LUMO+1 and HOMO-1 $\rightarrow$ LUMO transitions are used to fix the HOMO-1 and LUMO+1 levels relative to the HOMO level.

So far we have discussed how to locate the relative positions of the resonant peaks around the Fermi energy, but we still have to fix the Fermi energy itself. To reliably locate the Fermi energy relative to the corrected resonance peaks, we shift the whole spectrum, so that the HOMO energy equals $E_{\textrm{F}}-IP_{\textrm{mol}}$, where $IP_{\textrm{mol}}$ is the ionization potential of the molecule and \hbox{$E_{\textrm{F}}=5.1$\,eV} is the work function of gold\cite{AuWf}. The ionization potential $IP_{\textrm{mol}}$ is taken to be the difference between the total DFT energy of the neutral molecule, and that of the +$\vert e\vert$-charged molecule, using the equilibrium geometry of the neutral molecule in both cases. This definition assumes that on the timescale of tunnelling transport, the molecule has no time to relax to the equilibrium geometry of the charged state, which would shift the corresponding eigenenergies around the Fermi level. To compute $IP_{\textrm{mol}}$, we again used  NWChem with the same basis and exchange-correlation potential. For the transport calculation, the protecting groups were removed from the molecules, and for the TDDFT calculations, the dangling bonds were saturated with H atoms.

At this point, we have the corrected eigenenergies of the four states relative to the Fermi level, which we set to be the origin of energy. In case of non-zero off-diagonal $\kappa_{ij}$ elements, the correction of $H_M$ is carried out by first diagonalizing the $4\times 4$ molecular sub-matrix of $H_M$, then shifting the resulting eigenenergies to reproduce the excitations obtained from TDDFT, and finally transforming the matrix back to the original basis. Then the $4\times 4$ molecular sub-matrix of $H_M$ can be replaced by this corrected matrix to yield the new, corrected model Hamiltonian. This procedure ensures that we only shift the eigenenergies of the molecule and we do not alter the molecular eigenfunctions and therefore we preserve the original DFT ground-state density, which is expected to be accurate.

\section{Building blocks}\label{sec:blocks}

To understand how the different building blocks of these molecules influence transport properties, we examine transport through a related sequence of molecules, with varying levels of complexity. In this section and in section~\ref{sec:side-group}, the focus is on the qualitative trends contained in the uncorrected DFT-based transmission $T(E)$.

\begin{figure}[h]
\centering
  \igr[width=82mm]{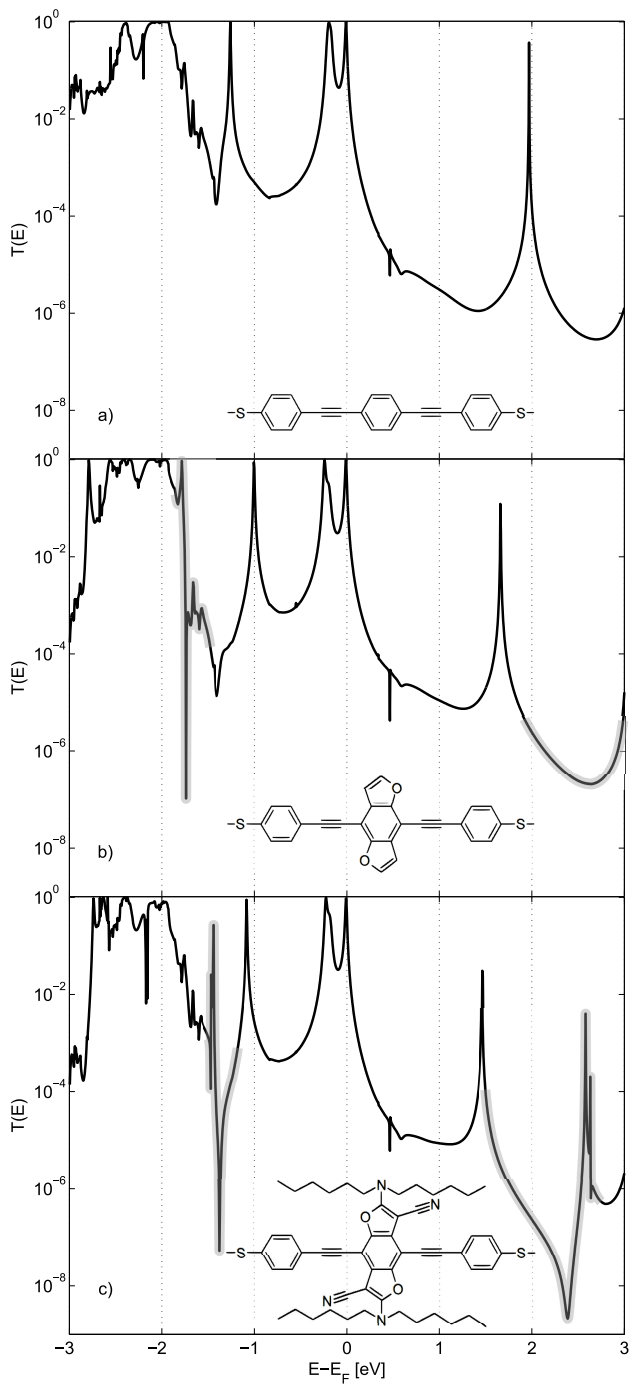}
  \caption{Zero-bias transmission $T(E)$ against electron energy $E$ for the triple-bonded BDF molecule {\bf 3} in stages of building blocks (uncorrected transport curves)}
    \label{fgr:BlockTran}
\end{figure}

The backbone of molecule {\bf 3} of Fig.~\ref{fgr:BDFstructure} is the widely-studied OPE (oligophenylene ethynylene) system shown in Fig.~\ref{fgr:BlockTran}(a) (inset).  For this structure, the zero-bias transmission $T(E)$ is shown in Fig.~\ref{fgr:BlockTran}(a), which reveals that the Fermi energy \hbox{($E_{\textrm{F}}\,=\,0\,$eV)} sits close to the HOMO resonance. The LUMO resonance possesses the shape of a Lorentzian curve, typical of Breit-Wigner transport resonances, whereas the HOMO forms a double-resonance with the close lying HOMO-1.

The next step towards molecule {\bf 3} is to replace the central phenyl ring with the BDF core, resulting in the transmission probability shown in (Fig.~\ref{fgr:BlockTran}(b)), whose HOMO-2 and the LUMO peaks are shifted closer to the Fermi energy. Around $2.5$\,eV, the local minimum is also decreased and close to $-2$\,eV, a sharp antiresonance now appears.

In Fig.~\ref{fgr:BlockTran}(c), which contains both cyano and amine groups, the transmission probability shows clear examples of Fano resonances\cite{FanoPRB}, where the line shape consists of a resonant peak followed or preceded by an anti-resonance. The addition of these groups lead to the appearance of these two resonances at $E=-1.5$\,eV, and at $E=2$\,eV. It is known that Fano resonances arise in molecular wires due to interference effects between a side group and the backbone of the molecule\cite{FanoPRB}. For molecule {\bf 3}, we can conclude that the difuran units with the cyano and amine groups are responsible. These additional groups have caused the Fano resonances to shift, however, the transport behaviour close to the Fermi energy is similar in all three cases shown in Fig.~\ref{fgr:BlockTran}.

\begin{figure}[h]
\centering
  \igr[width=86mm]{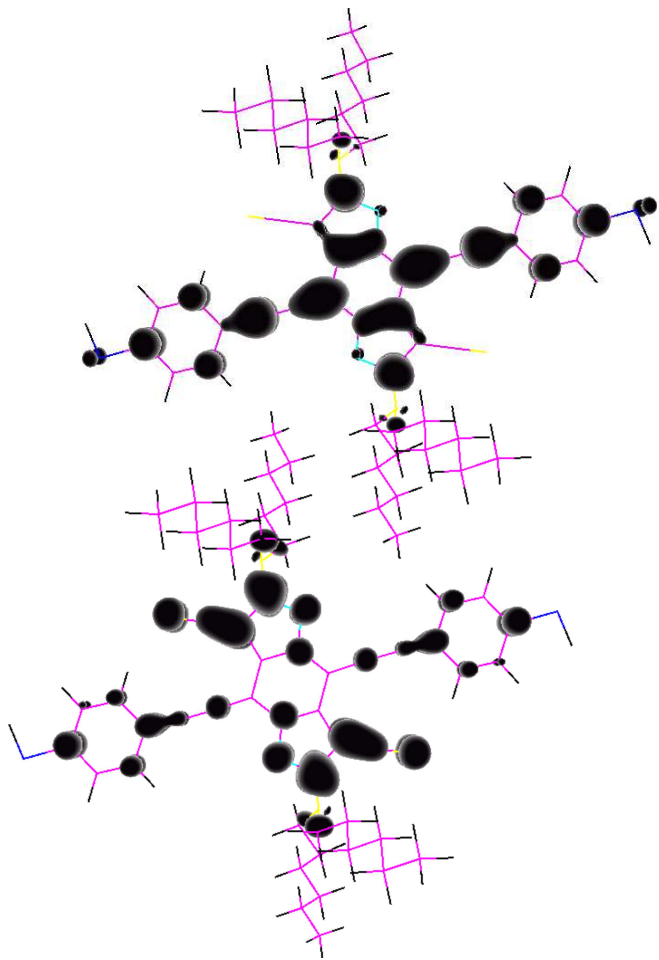}
  \caption{LUMO (top) and LUMO+1 (bottom) orbitals of the triple-bonded BDF molecule {\bf 3}.}
  \label{fgr:Orbs}
\end{figure}

Further understanding of the transport properties of molecule {\bf 3} can be extracted from the orbitals of the isolated molecule. Figure \ref{fgr:Orbs} shows the DFT calculated LUMO and LUMO+1 orbitals. The LUMO orbital is heavily weighted along the backbone of the molecule (the HOMO also shows similar behaviour) and this type of orbital typically produces a Breit-Wigner resonance. This contrasts with the LUMO+1 orbital, which is located mainly on the central BDF core. This orbital is responsible for the Fano resonance close to $2$\,eV in the transmission curve of Fig.~\ref{fgr:BlockTran}(c).

\section{Side group substitution}\label{sec:side-group}

The ability to control transport is one of the main aims of single molecule electronics. In the molecular junctions studied in this paper, the HOMO and LUMO resonances are produced by a delocalized orbitals, which are difficult to modify by an external stimulus. However Fano resonances are produced by a localized orbitals, which are more sensitive to environment changes or external fields. Here we carry out a simple theoretical experiment to illustrate how the position of a Fano resonance could be controlled through targeted design.
\begin{figure}[h]
 \centering
    \igr[width=86mm]{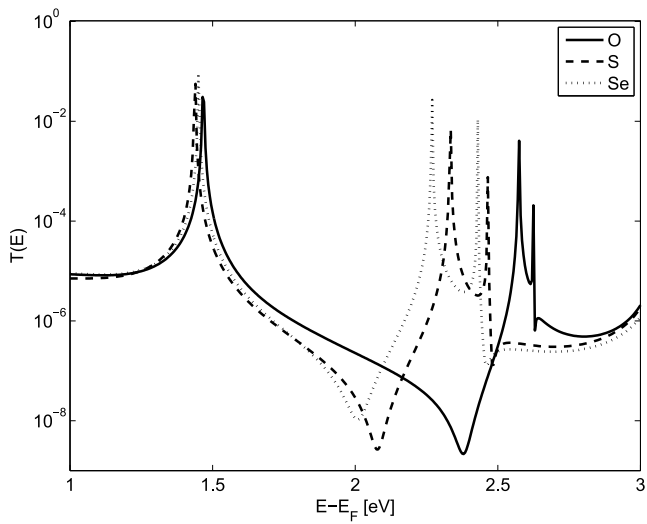}
    \caption{Transmission in the presence of different substituted atoms (uncorrected transport curves)}
    \label{fgr:SUB}
\end{figure}

The furan subunit of {\bf 3} can be changed by substituting different atoms in place of the oxygen atoms\cite{Fluor1}. In what follows, we replace the oxygens by the group-16 atoms, sulphur and selenium, and in each case, find the relaxed geometry of these new, triple-bonded molecules. As an example of the changes in geometry caused by this substitution, the calculated bond lengths between the substituted atom and the backbone are as follows: C-O bond = 1.38\,{\AA}, C-S = 1.76\,{\AA} and C-Se = 1.9\,{\AA}. Figure~\ref{fgr:SUB} shows the transmission curve through these three molecules, and focuses on the area close to the Fano resonance at 2\,eV. As expected, the HOMO resonance is largely unaffected, but we see that both the sulphur and selenium shift the Fano resonance closer to the Fermi energy. For sulphur, the antiresonance and associated resonance are shifted closer to the Fermi energy, with the antiresonance sitting 0.3\,eV below that of the oxygen.

\section{BDF series of molecular wires}\label{sec:BDFseries}

\begin{figure}[!h]
 \centering
    \igr[width=86mm]{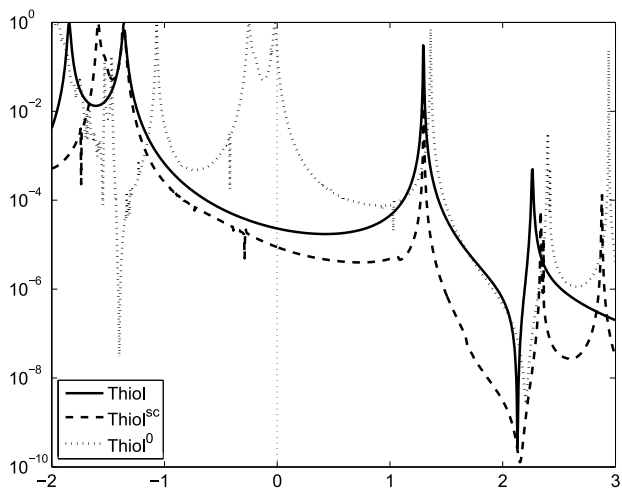}
    \caption{Zero-bias transmission $T(E)$ for the triple-bonded molecule with the correction in Sec.~\ref{sec:theory} (solid line), with the scissors correction method (dashed line) and without any correction (dotted line). The conductance values at the Fermi energy are $2.31\times10^{-5} G_0$, $8.78\times10^{-6} G_0$ and $2.19\times10^{-1} G_0$ respectively.}
\label{fgr:Thiol}
\end{figure}

So far, we have focussed on the uncorrected DFT-based transmission probability. Fig.~\ref{fgr:Thiol} presents results for the corrected transmission probability of molecule {\bf 3} (solid line) and the uncorrected transmission (dotted line). For comparison, the figure also includes results that would be obtained using an alternative correction method.

This alternative methods is the so-called scissors correction method\cite{Quek2007,SAINT} and yields the dashed line in Fig.~\ref{fgr:Thiol}. The scissors correction method is performed by diagonalizing the molecular sub-matrix of the full Hamiltonian, then shifting the eigenvalues below and above the Fermi energy such that the new HOMO-LUMO gap matches with the first singlet excitation value of the isolated molecule. Finally the diagonalized matrix is transformed back to the original basis to obtain the corrected full Hamiltonian. The main limitation with this method is that it applies a constant shift to all occupied levels of the discrete spectrum  relative to the unoccupied levels and does not allow occupied (or unoccupied) levels to shift relative to each other. In this sense, the method leading to $T_M(E)$ can be viewed as an extension of the scissors correction method\cite{Quek2007,SAINT}.

Fig.~\ref{fgr:Thiol} shows that the correction described in Sec.~\ref{sec:theory} produces a curve somewhat above the alternative method. But whichever correction method is used, the conductance is decreased by 3-4 orders of magnitude compared to that obtained from the uncorrected $T(E)$. This underlines the importance of correcting for the HOMO-LUMO gap, and the position of the Fermi energy inside the gap.

Next, we examine the effect of the bonding type along the backbone of the molecule by performing calculations on the series of molecules shown in Fig~\ref{fgr:BDFstructure}. Hereafter we only present corrected results. The single-bonded molecule {\bf 1} is the shortest, with a calculated length of 1.52\,nm, and the geometry is non-planar, with a twist angle of 44$^\circ$ between the planes of the individual rings. The lengths of the double-bonded and triple-bonded molecules are 1.99\,nm and 2.04\,nm respectively and both of these molecules are planar. We consider the trans-trans configuration of the double-bonded molecule {\bf 2}. (However there is little change in the conducting properties for the trans-cis configuration.) The corrected HOMO-LUMO gaps are presented in Table~\ref{tab:gap}.
\begin{table}[h!]
\caption{Theoretical values for the HOMO-LUMO gaps [eV]}
 \centering
 \begin{tabular}{c c c c}
 \hline\hline
  & single- & double- & triple- \\ [-0.5ex]
  & bonded BDF & bonded BDF & bonded BDF \\ [0.5ex]
  \hline
  HOMO-LUMO gap & & & \\
  by DFT without & 2.44 & 1.27 & 1.38 \\
  corrections & & & \\
  \hline
  1st singlet excitation & & & \\
  energy by TDDFT\footnote{The errors of all the TDDFT calculations are less than 1\,meV.} & 3.24 & 2.59 & ${}^{\ \ }$2.80\cite{Liu} \\ [1ex]
  \hline
  ionization potential & 6.18 & 5.95 & 6.46 \\ [1ex]
  \hline
  \end{tabular}
  \label{tab:gap}
\end{table}

In these molecules, the gap is underestimated by pure DFT with GGA by up to 50\%, but the general trend does not change. The single-bonded molecule {\bf 1} has the largest gap, which is a  consequence of its smallest size. It is less obvious why the triple-bonded has the second largest gap, but this can be also understood by considering the effect of bond-length alternation along the \hbox{$\pi$-conjugated} backbones~\cite{Kushmerick2002}, which is somewhat higher in the triple-bonded one than in the double-bonded one due to the stronger bonding.\footnote{Note that the gas phase TDDFT calculations cannot include the effect of the image charges in the leads, which tends to decrease the value of the HOMO-LUMO gap.}

\begin{figure}[h]
 \centering
    \igr[width=86mm]{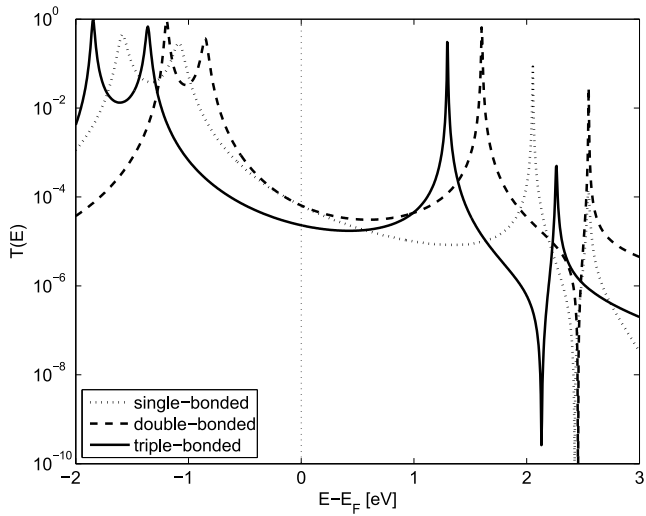}
    \caption{Zero-bias transmission $T_M(E)$ for single, double and triple-bonded molecules in Fig.~\ref{fgr:BDFstructure}. At the Fermi energy, $G_\textrm{Sb}=6.36\times10^{-5} G_0$, $G_\textrm{Db}=6.42\times10^{-5} G_0$ and $G_\textrm{Tb}=2.31\times10^{-5} G_0$.}
\label{fgr:series}
\end{figure}

The transmission coefficient $T_M(E)$ of the series of BDF molecules shown in Fig.~\ref{fgr:BDFstructure} can be seen in Fig.~\ref{fgr:series}, which shows that the conductance values at the Fermi energy are determined mainly by the position of the HOMO peak. Despite  having the largest HOMO-LUMO gap, and the twisted conformation\cite{Venkataraman2006}, the single-bonded molecule comes second after the double-bonded one. Without any corrections, the conductance values would be higher by four orders of magnitude, with the single-bonded having the lowest conductance: $G_\textrm{Sb}=1.86\times10^{-1} G_0$, $G_\textrm{Db}=4.52\times10^{-1} G_0$ and $G_\textrm{Tb}=2.19\times10^{-1} G_0$. Usually, in the case of thiol, a relatively high conductance is expected, as reported in various theoretical and experimental works with thiol and conjugated systems~\cite{Hong2012,Toni}. The high conductance is explained with strong thiol-gold coupling, since the sulphur atom tends to bind to multiple gold atoms preferentially in hollow position. However in the presented geometry the sulphur atoms binds to a single apex gold atom, thus resulting rather an estimated minimum for the conductance. The apex bonding geometry also accounts for the narrow resonance peaks.

\section{Thermoelectric properties}\label{sec:thermo}

We now turn to the room-temperature thermoelectric properties of the above molecules, which can be obtained from weighted integrals of $T(E)$ around the Fermi energy~\cite{Finch2009}.
\begin{table}[h!]
\caption{Conductance ($G$), thermal conductance ($\kappa$) and thermopower ($S$) for the single-, double- and triple-bonded molecules at room temperature (300K)}
 \centering
 \begin{tabular}{l c c c}
 \hline\hline
  & $G$ & $\kappa$ & $S$ \\ [-0.5ex]
  & [nS]; [$G_0$] & [keV/s/K] & [$\mu$V/K] \\ [0.5ex]
  \hline
  1 - Single-bonded\ & 4.99; $6.44\times 10^{-5}$ & 233 & 21.5 \\
  \hline
  2 - Double-bonded\ & 5.05; $6.51\times 10^{-5}$ & 238 & 19.7 \\
  \hline
  3 - Triple-bonded\ & 1.80; $2.32\times 10^{-5}$ & 83.3 & 9.62 \\
  \hline
  \end{tabular}
  \label{tab:thermo}
\end{table}
The resulting electrical conductance $G$, the thermal conductance $\kappa$ and the thermopower $S$ are shown in Table~\ref{tab:thermo}. The thermopower or Seebeck coefficient $S$ of a material or of a nanojunction is defined as $$S=-\frac{\Delta V}{\Delta T}\, ,$$ where $\Delta V$ is the voltage difference between the two ends of the junction when a temperature difference $\Delta T$ is established between them. This quantity controls the direct conversion of heat into electric energy at a molecular scale. If $T(E)$ varies only slowly with energy on the scale of $k_BT$, where $T$ is the temperature, $k_B$ Boltzmann's constant, then the exact expression~\cite{Finch2009} for $S$ simplifies to $$S=-\frac{\pi^2 k_B^2 T}{3e}\frac{\partial \ln T(E)}{\partial E}$$ evaluated at $E=E_F$, and $e$ is the magnitude of the electronic charge. For example, this means that if the LUMO resonance is closer to the Fermi energy than the HOMO, such that the slope of $\ln T(E)$ at $E=E_F$ is positive, then $S$ will be negative.

Table~\ref{tab:thermo} shows that the thermopower of molecules \textbf{1} and \textbf{2} are comparable with the highest values of $S$ measured to date for single molecules. For example, recently-measured values of $S$ at room temperature include $8.7$, $12.9$ and $14.2$ $\mu$V/K for 1,4-benzenedithiol (BDT), 4,4'-dibenzenedithiol, and 4,4''-tribenzenedithiol in contact with gold respectively~\cite{A1}, $-1.3$ to $8.3$ $\mu$V/K for the benzene-based series of benzene-dithiol (BDT), 2,5-dimethyl-1,4-benzenedithiol (BDT2Me), 2,3,5,6-tetrachloro-1,4-benzenedithiol (BDT4Cl), 2,3,5,6-tetraflouro-1,4-benzenedithiol (BDT4F) and BDCN~\cite{A2,A3}, $7.7$ to $15.9$ $\mu$V/K for the series BDT, DBDT, TBDT and DMTBDT~\cite{A4}, $-12.3$ to $13.0$ for a series of amine-Au and pyridine-Au linked molecules~\cite{A5}.

The thermopower of molecules \textbf{1} and \textbf{2} exceed all of these values. Only fullerene-based single-molecule junctions~\cite{A6,A7,A8} with $S$ ranging from $-8.9$ to $-33.1$ $\mu$V/K have been measured to have higher values.

\section{Conclusions}

The conductance of a series of short molecular wires containing BDF cores has been studied to assess their performance as components in molecular devices. The appearance of Fano resonances has been investigated and we have demonstrated that the positions of Fano resonance can be controlled via the electronegativity of the atoms of the side groups attached to the backbone.  We have also made a comparison between the single-, double- and triple-bonded BDF molecules with thiol anchor groups contacting apex gold atoms.

The room-temperature thermoelectric parameters of the given molecules have been evaluated. Table~\ref{tab:thermo} shows that the largest thermopower is predicted for molecule {\bf 1}. This value is comparable with the highest measured values for single-molecule thermopower reported to date. The thermopower is approximately proportional to the slope of $\ln T_M(E)$ at $E=E_F$ and therefore the high value of $S$ for molecule {\bf 1} is a reflection of the high slope of the corresponding curve in Fig~\ref{fgr:series}. This high slope is a consequence of the suppression of the LUMO resonance for this molecule, compared with {\bf 2} and {\bf 3}. For all molecules considered, the Fano resonances are not located close to $E_F$ and play only a minor role in determining thermoelectric coefficients.

\acknowledgments

The authors wish to acknowledge support from the Marie-Curie ITNs MOLESCO and NanoCTM and from Engineering and Physical Sciences Research Council (UK) (EPSRC). Special thanks to Silvio Decurtins, Shi-Xia Liu and Toni Fr\"olich who provided us with valuable discussions and experimental data.


%

\end{document}